\documentclass[a4paper]{spie}  
\usepackage[]{graphicx}
\usepackage{amssymb,amstext}
\usepackage{cite}
\usepackage{hyperref}

\def\sun{\hbox{$\odot$}}

\def\arcsec{\hbox{$^{\prime\prime}$}}

\title{The Science Case for the Planet Formation Imager (PFI)\footnote{\small~Copyright 2014 Society of Photo-Optical Instrumentation Engineers. One print or electronic copy may be made for personal use only. Systematic reproduction and distribution, duplication of any material in this paper for a fee or for commercial purposes, or modification of the content of the paper are prohibited. DOI abstract: http://dx.doi.org/10.1117/12.2055544}}

\author{Stefan Kraus\supit{a}, 
John Monnier\supit{b},
Tim Harries\supit{a},
Ruobing Dong\supit{c},
Matthew Bate\supit{a},\\
Barbara Whitney\supit{d},
Zhaohuan Zhu\supit{c},
David Buscher\supit{e},
Jean-Philippe Berger\supit{f},
Chris Haniff\supit{d},
Mike Ireland\supit{g},
Lucas Labadie\supit{h},
Sylvestre Lacour\supit{i},
Romain Petrov\supit{j},
Steve Ridgway\supit{k},\\
Jean Surdej\supit{l},
Theo ten Brummelaar\supit{m},
Peter Tuthill\supit{n},
Gerard van Belle\supit{o}
\skiplinehalf
\supit{a}University of Exeter, School of Physics, Stocker Road, Exeter, EX4 4QL, UK;
\supit{b}Department of Astronomy, University of Michigan, 500 Church St., Ann Arbor, MI 48109, USA; 
\supit{c}Department of Astrophysical Sciences, Princeton University, Princeton, NJ 08544, USA; 
\supit{d}Astronomy Department, University of Wisconsin-Madison, 475 N. Charter St., Madison, WI 53706, USA;
\supit{e}Cavendish Laboratory, JJ Thomson Avenue, Cambridge, CB3 0HE, UK;
\supit{f}European Southern Observatory, 85748, Garching by M\"{u}nchen, Germany;
\supit{g}Research School of Astronomy \& Astrophysics, Australian National University, Canberra ACT 2611, Australia;\\
\supit{h}I.~Physikalisches Institut, Universit\"{a}t zu K\"{o}ln, Z\"{u}lpicher Strasse 77, 50937 Cologne, Germany;
\supit{i}Laboratoire d'Astrophysique de Grenoble, UMR 5571 Universit\'{e} Joseph Fourier/CNRS, BP 53, 38041 Grenoble Cedex 9, France;
\supit{j}Laboratoire Lagrange, UMR7293, Universit\'{e} de Nice Sophia-Antipolis, CNRS, Observatoire de la C\^{o}te d’Azur, Bd. de l’Observatoire, 06304 Nice, France;
\supit{k}National Optical Astronomy Observatory, P.O. Box 26732, Tucson, AZ 85726-6732, USA;
\supit{l}lDepartment of Astrophysics, Geophysics and Oceanography (AGO), AEOS Group, Li\`ege University, All\'ee du 6 Ao\^ut 17, 4000 Li\`ege, Belgium;
\supit{m}The CHARA Array, Mount Wilson Observatory, Mount Wilson, CA 91023, USA;\\
\supit{n}Sydney Institute for Astronomy, School of Physics, University of Sydney, NSW 2006, Australia;
\supit{o}Lowell Observatory, 1400 West Mars Hill Road, Flagstaff, AZ 86001, USA
}

\authorinfo{Further author information: \\
Send correspondence to S.K., E-mail: skraus@astro.ex.ac.uk, Telephone: +44 1392 724125}

\begin{document} 
  \maketitle 

\begin{abstract}

Among the most fascinating and hotly-debated areas in contemporary astrophysics 
are the means by which planetary systems are assembled from the large rotating 
disks of gas and dust which attend a stellar birth. Although important work has already
been, and is still being done both in theory and observation, a full understanding of the physics 
of planet formation can only be achieved by opening observational windows able 
to directly witness the process in action. The key requirement is then to probe 
planet-forming systems at the natural spatial scales over which material is being assembled.  
By definition, this is the so-called Hill Sphere which delineates the region of influence of 
a gravitating body within its surrounding environment. 

The Planet Formation Imager project (PFI; \url{http://www.planetformationimager.org}) 
has crystallized around this challenging goal: 
to deliver resolved images of Hill-Sphere-sized 
structures within candidate planet-hosting disks in the nearest star-forming regions. 

In this contribution we outline the primary science case of PFI. For this purpose, 
we briefly review our knowledge about the planet-formation process and discuss recent 
observational results that have been obtained on the class of transition disks. 
Spectro-photometric and multi-wavelength interferometric studies of these systems 
revealed the presence of extended gaps and complex density inhomogeneities that might 
be triggered by orbiting planets. We present detailed 3-D radiation-hydrodynamic 
simulations of disks with single and multiple embedded planets, from which we 
compute synthetic images at near-infrared, mid-infrared, far-infrared, 
and sub-millimeter wavelengths, enabling a direct comparison 
of the signatures that are detectable with PFI and complementary facilities such as ALMA. 
From these simulations, we derive some preliminary specifications that will guide 
the array design and technology roadmap of the facility.

\end{abstract}



\section{INTRODUCTION}
\label{sec:intro}

From the first theories on the origin of our solar system in the 18th century 
(by Emanuel Swedenborg, Immanuel Kant, and Pierre-Simon Laplace) 
up to the late 1990's, our understanding of the planet formation process has been
guided solely by the characteristics observed in our own planetary system.
This resulted in the classical ``core-accretion theory'',
where the planets are assembled already at their final location through dust coagulation 
and agglomeration processes (e.g.\ Pollack et al.\cite{pol96}).
The discovery of the first extrasolar planet around a main-sequence star 
by Mayor \& Queloz\cite{may95} and the findings of the subsequent exoplanet surveys 
changed this view dramatically and revealed a surprising diversity in planetary
system architecture.  Many of the detected systems exhibit Jupiter-mass planets
that orbit their host star at separations of less than 0.1\,astronomical units (``Hot Jupiter'') 
or planets with masses of $\sim 10$ Earth masses (``Super-Earth'').
Neither of these planet populations is observed in our solar system, which raises 
the important question of what determines the diversity of these systems
and the properties of the assembled planets.

In order to account for this menagerie of planetary systems, various planet migration mechanisms 
and alternative formation scenarios (such as the gravitational instability model) have been proposed.  
Accordingly, in state-of-the-art population synthesis models the final system architecture 
depends on a plethora of parameters, describing the initial conditions of the protoplanetary disk, 
the planetesimal formation mode, the interaction between the planet and disk (type I/II migration),
the presence of migration traps (deadzones, disk truncation, ...), 
planet-planet scattering events (resonances, planet ejection, ...), 
environmental factors affecting the disk evolution, and 
finally scattering caused by the planetesimal disk.
These new ideas make planet formation one of the most vibrant and 
active research fields, but reveal also the complexity that results from the large 
number of involved processes.
It is becoming increasingly clear that new observational constraints are needed, 
both to determine the initial conditions of the planet formation process and to identify the 
dominant mechanisms that govern the assembly and orbital evolution of planetary systems.

With the Planet Formation Imager (PFI) project, we argue that these urgently needed
observational constraints could be obtained with a new high-angular resolution imaging facility
that would be optimized to probe these processes on the natural spatial scale where 
planet formation is taking place.  This natural spatial scale is the ``Hill Sphere'', which 
defines the gravitational sphere of influence of the forming planet.
The Hill Sphere of a Jupiter-mass planet at the location of Jupiter in our solar system 
is 0.35\,astronomical units (au, for $r=5.2$\,au) and 0.07\,au for a Jupiter-mass planet at $r=1$\,au.

From these ambitious goals we launch our efforts and have started to bring together a
science working group (SWG) that will be charged with answering fundamental questions, like:
What wavelengths and spatial scales will be key for PFI? 
What different aspects of planet formation can we learn from scattered light, 
thermal IR, or mm-wave imaging?  What are the optimal diagnostic lines to determine 
the physical conditions and kinematics in the circumplanetary accretion disk? 
Which nearby star-forming regions are the most important to survey? 
Can PFI also detect the panoply of warm exoplanets expected to be around most young stars? 

The SWG will include observers and theoreticians to address these and other open questions.
We will conduct detailed simulations and interact with the 
technical working group (TWG) in order to develop a roadmap for implementing PFI 
at the budget of a mid-scale international facility project.

In the following chapters we review some of the recent advancements in high-angular resolution
observations on planet-forming disks (Sect.~\ref{sec:context}) and
introduce the PFI project (Sect.~\ref{sec:pfi}) as well as our working plan 
(Sect.~\ref{sec:swg}).  We conclude in Sect.~\ref{sec:conclusions}.
Further details about our organizational structure and technological considerations 
are given in accompanying articles by Monnier et al., Ireland et al., and Buscher et al.\  in this volume.

\section{STATE-OF-THE-ART IN HIGH ANGULAR RESOLUTION
STUDIES ON PLANET FORMATION}
\label{sec:context}

Planet formation studies are experiencing a tremendous 
level of activity, driven both by theoretical advancements
and the fundamentally new observational capabilities that are 
provided by the new high-angular resolution imaging capabilities of ALMA
and adaptive optics systems like Gemini/PFI and VLT/SPHERE.

An important discovery of the last decade has been the class of
transitional and pre-transitional disks (Calvet et al.\cite{cal02}, Espaillat et al.\cite{esp07}).
These objects exhibit a strong mid-infrared (MIR) excess, but have a 
significantly reduced near-infrared (NIR) excess compared to T\,Tauri or 
Herbig\,Ae/Be disks (Espaillat et al.\cite{esp08}). This reduced NIR excess emission 
indicates that the innermost disk regions contain only optically thin gas and dust 
(transitional disks) or exhibit an extended gap, which separates the optically thick 
inner disk from the outer disk (pre-transitional disks). 

Coronagraphic observations using the Subaru/HiCIAO and VLT/NACO instruments revealed 
intriguing spiral arm-like structures in the outer regions of several transitional and
pre-transitional disks (e.g.\ Muto et al.\cite{mut12}, Quanz et al.\cite{qua13b})
that might be triggered by embedded planets.
Structures on smaller spatial scales were probed using interferometry at multiple wavelengths.
At sub-millimeter wavelengths the SMA, CARMA, ALMA, and the
Plateau de Bure Interferometer detected central density depressions on scales of tens of au,
consistent with an extended inner disk hole or a gapped disk structure
(e.g.\ Isella et al.\cite{ise10}, Andrews et al.\cite{and11}, Casassus et al.\cite{cas13}).
Near- and mid-Infrared interferometry with the VLTI/AMBER, MIDI, and PIONIER instruments
resolved the distribution of material on (sub-)au scales and
characterized the conditions inside the gap.
Observations on HD\,100546, T\,Cha, and V1247\,Ori showed that the
near-infrared (1-2\,$\mu$m) emission is dominated by the emission of
a narrow inner disk component located near the dust sublimation radius and
smaller contributions from scattered light (Oloffson et al.\cite{olo13}) and 
optically thin dust emission from within the gap (Kraus et al.\cite{kra13}).
The mid-infrared regime (N-band) is sensitive to 
a wider range of dust temperatures and stellocentric radii and
includes emission contributions from the inner disk, the disk gap, and the outer disk.
The gaps of HD\,100546 and T\,Cha were found to be
highly depleted of (sub-)$\mu$m-sized dust grains,
with no significant near-/mid-infrared emission 
(Benisty et al.\cite{ben10}, Oloffson et al.\cite{olo13}),
while the disk gap of V1247\,Ori is filled with optically thin 
dust material (Kraus et al.\cite{kra13}).
TW\,Hya appears to be in a particularly late stage
of disk clearing, with a very extended dust-depleted
inner hole that extends out from $\sim 0.3$\,au
(Menu et al.\cite{men14}) and contains large, settled dust grains.

A particularly intriguing finding obtained with interferometry 
on transitional disks is the detection of non-zero phase signals, 
which indicates the presence of strong asymmetries in the 
inner, au-scale disk regions.  
Some detections are consistent with low-mass companions, 
such as the possibly planetary-mass body around the 
transitional disk of LkCa\,15 (Kraus \& Ireland\cite{kra12}).
In other cases, the asymmetries can be attributed to emission contributions 
from a vertically extended disk seen under intermediate inclination angle
(e.g.\ T\,Cha and FL\,Cha; Olofsson et al.\cite{olo13} and Cieza et al.\cite{cie13}).
Another interesting case is V1247\,Ori, where
the asymmetries seem to trace complex, radially extended 
disk structures (Kraus et al.\cite{kra13}) that might be caused by
the dynamical interaction of the (yet undiscovererd) gap-opening body/bodies 
with the disk material.

\section{THE PFI PROJECT}
\label{sec:pfi}

The latest observational and theoretical studies suggest that planet formation
is a highly complex and multi-facetted process, where planet-induced
dynamical processes are accompanied by complex changes in the dust mineralogy.
High-angular resolution instruments both at infrared and sub-milllimeter
wavelengths are just starting to obtain a first glimpse of this complexity,
but are not able to properly resolve and characterize these intruiging structures.

Further advancements in our understanding of the planet formation process
can be expected in the coming years, for instance from the fully-operational ALMA array, 
the VLTI 4-telescope mid-infrared interferometric instrument MATISSE 
(Lopez et al.\cite{lop06}), and the upcoming generation of extremely large telescopes (ELTs). 
However, in the relevant infrared/sub-mm wavelength regime, these facilities will 
still be limited to angular scales 
of {$\sim 0.01$\arcsec} or 1.5\,au for the most nearby star forming regions,
which is $\sim 30\times$ larger than the
Hill Sphere of a Jupiter-like object and $\sim 100\times$ 
larger than the circumplanetary accretion disk.
A facility with even higher resolution will be essential to
probe the processes on the natural scale, where planet formation is happening.

As a consequence of this realisation, the PFI project was born in 
late 2013 and an international consortium has rapidly emerged 
to develop realistic project goals, with participants drawn mainly from 
the high resolution astronomical imaging community. 
Scientists from more than a dozen different institutes in six countries 
are presently on the project launch committee.
The project executives have been elected in February 2014,
including the Project Director John Monnier (University of Michigan, USA), 
Project Scientist Stefan Kraus (University of Exeter, UK), and 
Project Architect David Buscher (University of Cambridge, UK). 
During the next 1-2 years, we will develop and prioritize the science goals 
and consider all technologies and facility architectures that might be 
capable of achieving our science objectives, including visible, 
thermal infrared, and mm-wave imaging from the ground and from space.


\section{THE PFI SCIENCE WORKING GROUP}
\label{sec:swg}

The SWG will be lead by the Project Scientist and will be 
reponsible for developing and maintaining the top-level science requirements of PFI.
It will be charged with investigating the signatures of planet formation 
at different stages of disk evolution, assessing the observability of the 
protoplanetary bodies, and to determine how PFI could significantly advance 
our understanding of the architecture and the potential habitability of planetary systems. 

Our results will be published in white papers and in 
refereed journals when appropriate. 
Scientists from the star formation, planet formation, planetary science, 
and exoplanet community are kindly invited to contribute to our efforts. 
In the following sections, we list the main topics that we 
plan to investigate within our SWG sub-groups.

\subsection{Protoplanetary Disk Structure \& Disk Physics}

Protoplanetary disks set the initial conditions of planet formation and 
determine the later dynamical evolution of the forming planetary system.
Fundamental aspects about disk structure and disk evolution are still poorly understood, 
reflecting the fact that most of our knowledge about protoplanetary disk structure 
has been derived by fitting spatially unresolved constraints, 
such as line profiles or the spectral energy distribution.

Once it is in full operation, ALMA will be able to provide further insights 
on the density structure of protoplanetary disks.  However, with an angular resolution 
of $\ge 5$\,milliarcsecond, ALMA will not be able to probe the disk regions in 
the inner-most au, where substantially different processes are believed to be at work 
than in the more extended disk.  For instance, the disk is believed to be truncated
at the co-rotation radius at a few stellar radii and it has been proposed that this
truncation might be responsible for the pile-up of the Hot Jupiter population (Lin et al.\cite{lin96}).

Designing PFI to operate at mid-infrared wavelengths would 
allow us to make use of complementary aspects with ALMA.
The sub-millimeter regime probed by ALMA is most sensitive to the thermal
emission of cold, mm-sized dust grains located in the outer disk and in the disk midplane.
The mid-infrared emission, on the other hand, traces small $\mu$m-sized dust grains
and the warm dust located in the disk surface layer.
Hydrodynamic simulations predict that the dust distribution can differ significantly
for different dust populations.  For instance, theories of dust filtration predict that 
large (mm-sized) dust grains are held back in the outer disk and accumulate in a narrow ring 
outside of the gap, while small ($\mu$m-sized) grains filter through the gap (Zhu et al.\cite{zhu12}). 
Spatial variations in the distribution of dust grain populations were also found by the recent
study on the transitional disk Oph\,IRS48 (van der Marel et al.\cite{van13}), where 
the small dust grains are distributed rather homogeneously throughout the disks, 
while the large dust grains are confined towards one part of the disk.
These differences in the distribution of large and small grains were modelled with 
a dust trap that might be triggered by an undetected planetary-mass companion. 
Being one of the most nearby (120 pc) transitional disks with a very extended 
inner hole ($\sim 40$\,au), it was possible to resolve the distribution of 
the $\mu$m-sized grains for Oph IRS48 with conventional VLT/VISIR imaging 
observations.  However, it is clear that higher resolution will be needed to study these
spatial dust variations in details, in particular for lower-luminosity objects
and in earlier evolutionary stages.

Interferometric observations in the mid-infrared wavelength regime (e.g.\ N-band)
also give access to various strong spectral features such as the 10\,$\mu$m Silicate feature 
and several hydrocarbon-related (PAH) features.
Spatially and spectrally resolved investigations with the VLTI/MIDI instrument 
revealed radial gradients in the dust minerology and found that the crystallinity 
is higher in the inner few au of the disk, supporting theories of grain growth 
(van Boekel et al.\cite{van04}). 
The hydrocarbon-related emission is located in the gap region 
(V1247\,Ori: Kraus et al.\cite{kra14}) and in the outer disk (Olofsson et al.\cite{olo13}).

\subsection{Planet Formation Signatures in Pre-Main-Sequence Disks}
\label{sec:pfsignatures}

\begin{figure}[p]
  \centering
  \includegraphics[height=21cm]{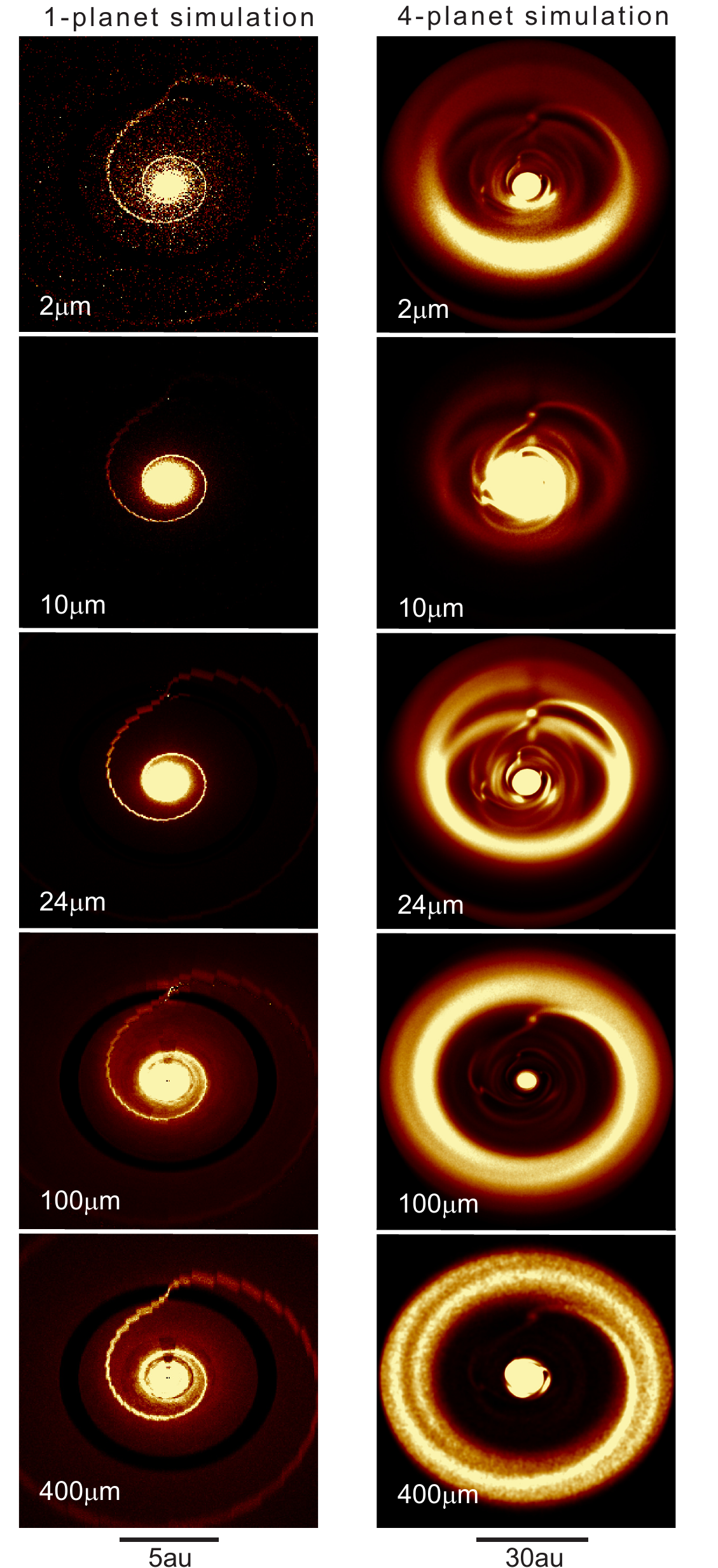}
  \caption{
    Radiative transfer simulation of planet-forming disks with
    a single planet (left; simulation from Harries) and 4 planetary bodies (right; simulation from Dong, Whitney \& Zhu), respectively.
    For details about the simulation set up, we refer the reader to
    Sect.~\ref{sec:pfsignatures}.
  }
  \label{fig:sim}
\end{figure}

Once planets have formed in the disks, they dynamically sculpt their environment, 
for instance by opening tidally-cleared gaps or triggering spiral arms and disk warps.
The aim of this sub-group is to make quantitative predictions for the 
structures that we can expect to detect with PFI, both in the dust continuum emission
and in spectral lines.  Below, we show radiative transfer simulations
of structures that we might expect to detect with PFI.  All simulations were computed
for an inclination angle of $30^{\circ}$ (i.e.\ closer to a face-on viewing angle) 
and a central star with an effective temperature of 4000\,K, a mass of 0.5\,M$_{\sun}$, 
a stellar radius of 2\,R$_{\sun}$, and solar abundance.\newline

\noindent{\it 1-Planet Radiative Transfer Simulation:} 

Our first simulation represents a relatively simple case of a protoplanetary disk
with a single Jupiter-mass planet located at
a separation of 5\,au from the central star.
This scenario could be observed during the T\,Tauri phase at an age around $0.5$\,Myr.
The radiative transfer computation was performed using the TORUS radiative transfer 
code by Tim Harries\cite{har00}.
The planet cleared a gap with a width of 1\,au and triggered
a spiral wake that was parameterised using the analytical description
by Ogilvie and Lubow\cite{ogi02}, closely matching the results of 
hydrodynamic simulations. 
The density perturbation is $10^4\,(M_{p}/M_{\star})$ (where
$M_{p}$ is the mass of the planet and $M_{\star}$ the stellar mass), 
or a factor 10 in density over the unperturbed disk density. 
We computed the images for five wavelengths that might be of interest
for PFI, including the near-infrared (K-band around 2\,$\mu$m),
mid-infrared (N- and Q-band around 10 and 24\,$\mu$m),
far-infrared (100\,$\mu$m), and sub-mm regime (400\,$\mu$m).
Containing only a single planet, the images exhibit the minimum amount
of complexity that we should expect to detect with PFI.\newline

\noindent{\it 4-Planet Radiative Transfer Simulation:} 

Our second set of simulations represents a later stage
of disk evolution, resembling the conditions that are expected during the 
pre-transitional disk phase, where an extended gap region has
been cleared by multiple planets.  The simulated images cover a 
field-of-view of 80\,au (side-to-side) at a pixel size of 0.1\,au.\newline
The simulation has been conducted using the HOCHUNK3D radiative transfer
code by Barbara Whitney et al.\cite{whi13}.  
The density profile for these simulations was computed using a
2-D hydrodynamic simulation by Zhaohuan Zhu\cite{zhu11},
which incorporates four Jupiter-mass planets located at
separations of 5, 7.5, 12.5, and 20\,au.  
The vertical disk scale height was computed based on hydrostatic equilibrium,
where we assume two dust populations:
A small dust population (with grain sizes up to $\sim 1\,\mu$m) that contributes 
10\% of the total dust mass and is well mixed with the gas component.
A bigger dust grain population (with sizes extending to 1\,mm) is collapsed to
20\% of the gas scale height, mimicking the effect of dust grain settling (see Dong et al.\cite{don12}). 

A qualitative inspection of these continuum images suggests that the mid-infrared regime 
(e.g.\ L-/M-/N-band between 3\,$\mu$m and 13\,$\mu$m)
might constitute a good wavelength regime to image the planet-related disk signatures
in the inner few au.
This wavelength regime is sensitive to the thermal emission of dust grains in the $\sim 1000-300$\,K
range, matching the temperature of hot dust in the disk surface layer in the terrestrial planet-forming region 
at a few au, as well as the expected temperature of the circumplanetary accretion disk (see below).
Shorter wavelengths (e.g.\ K-band around 2\,$\mu$m) contain significant contributions
from scattered light, while longer wavelengths (e.g.\ FIR at sub-mm) are dominated
by the emission of cold material in the outer disk.
Further work by the SWG and TWG will need to confirm this assessement
and identify the wavelength band(s) that provide the optimal balance between 
the brightness of the expected features and the technically achievable image contrast/imaging fidelity.

Beside the continuum tracers, we will also simulate the expected signatures 
in different line tracers.  Some interesting line tracers in the mid-infrared part of the spectrum 
could include the CO fundamental lines (4.7\,$\mu$m), CH$_3$OH ices (3.5\,$\mu$m), 
NH$_3$ ice (3.0\,$\mu$m), and C-H nanodiamonds (3.4-3.5\,$\mu$m).
Recording spectrally dispersed data in the line tracers will allow us to
construct maps in different velocity channels, constraining the kinematics of the gas
in the accretion streams and the circumplanetary accretion disk.

By imaging the disks at multiple epochs, we will also be able to study the 
temporal evolution of the planet-induced disk structures and to trace the 
orbital motion of the embedded protoplanets.  
At its' unprecedented, sub-milliarcsecond resolution, PFI will likely be
able to detect such structural changes on timescales of just a few months,
directly revealing the ongoing, highly dynamical planet formation processes.

\subsection{Protoplanet Detection}

\begin{figure}[t]
  \centering
  \includegraphics[height=10cm,angle=270]{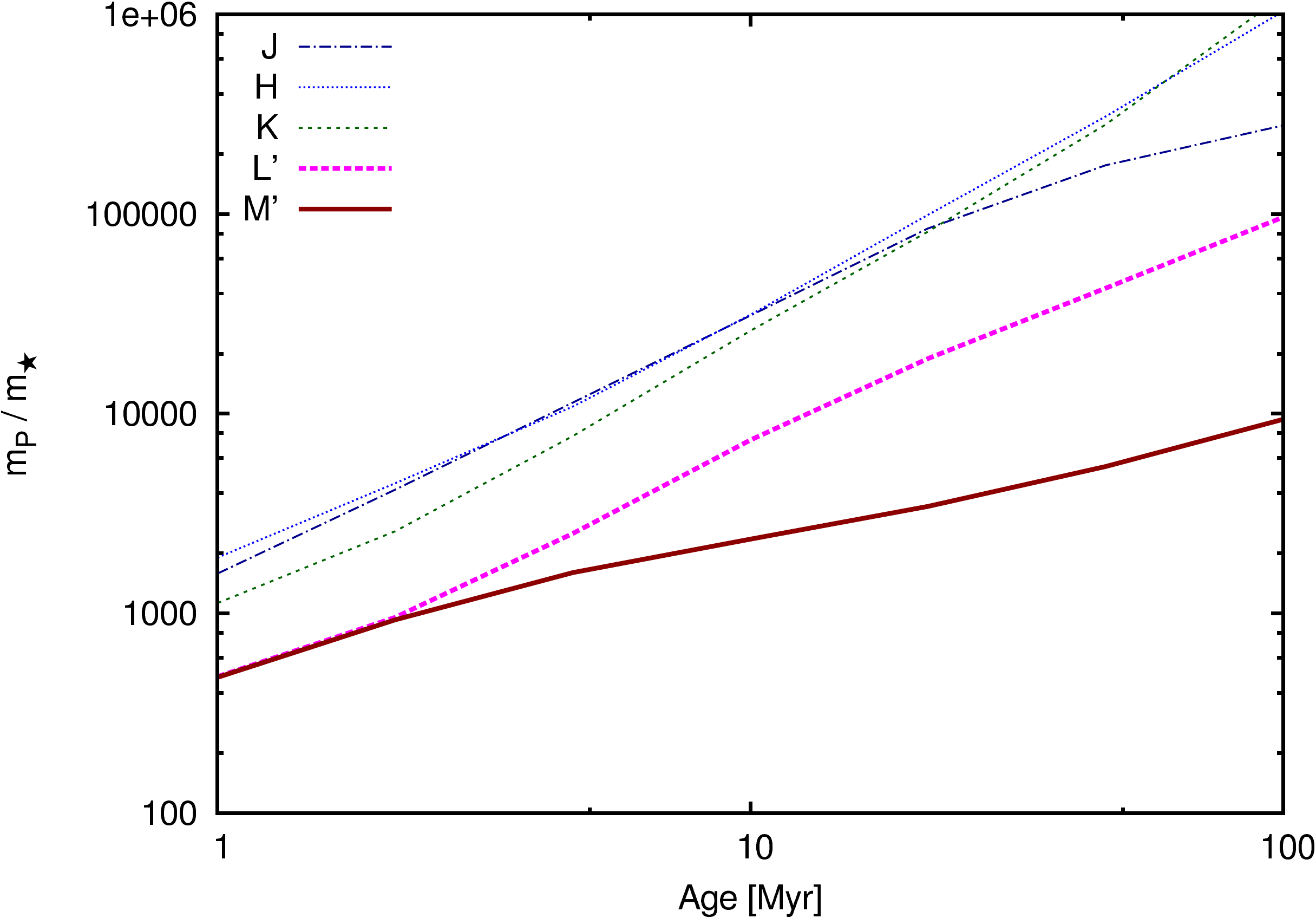}
  \vspace{3mm}
  \caption{
    Predicted planet/star contrast for a $4M_{\rm Jup}$ protoplanet,
    plotted as function of age.
    The computation is based on the atmosphere models by Baraffe et al.\cite{bar98,bar02}
    and on the ``Hot Start'' evolutionary tracks by Fortney et al.\cite{for08}
    and has been conducted for five spectral bands.
  }
  \label{fig:planetcontrast}
\end{figure}

\begin{figure}[t]
  \centering
  \includegraphics[height=10cm]{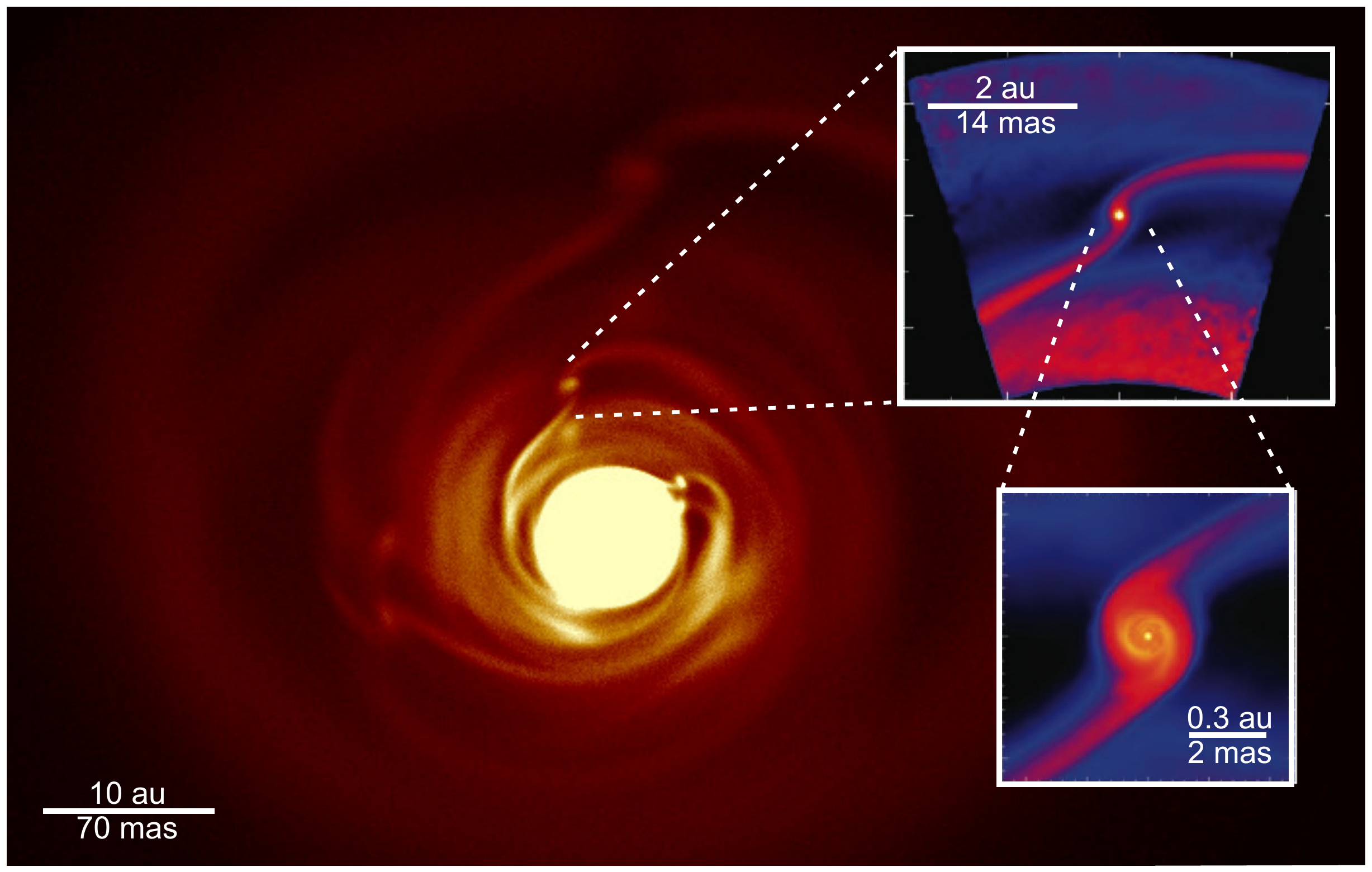}
  \caption{
    Mosaic of hydrodynamics simulations of a protoplanetary disk with four embedded planets
    (background), the gap that is opened by one of the planets (top-left panel), 
    and of the circumplanetary accretion disk (bottom-left panel), illustrating the need
    to probe a wide range of spatial scales in order to study the mechanisms that
    are at work during planet formation.
    The background image is the 10\,$\mu$m model image of the four-planet simulation
    by Dong, Whitney \& Zhu described in Sect.~\ref{sec:pfsignatures}, while
    the insets show surface density profiles from simulations by Ayliffe \& Bate\cite{ayl09}.
    Besides the physical scales (in au) we give the angular scale on the sky 
    for a distance of 140\,pc (in mas), which corresponds to the distance of the most
    nearby star forming regions (e.g.\ Taurus).
  }
  \label{fig:sketch}
\end{figure}

The simulations discussed above show the accretion streams towards the
forming planets, but do not include the emission from the protoplanet itself.
For young planets, the brightness of the viscously heated circumplanetary disk
can significantly exceed the thermal emission of the surrounding dust.
In the age range between $10^{6}$ and $10^{7}$ years, the brightness ratio between the 
central star and the protoplanet is predicted to reach $10^{3...4}$
in the mid-infrared wavelength regime (e.g.\ L-band, Figure~\ref{fig:planetcontrast}), 
which is much more favourable than during the main-sequence phase, when the
corresponding contrast is typically $10^{7...8}$.

Some individual protoplanet candidates have already also been detected,
including T\,Cha (Huelamo et al.\cite{hue11}) and LkCa\,15 (Kraus \& Ireland\cite{kra12}).
These two detections have been achieved with the aperture masking interferometry technique,
which offers an efficient way to reach a good phase accuracy ($\sim 0.1^{\circ}$)
near the diffraction-limit of large single-dish telescope.
However, the detections have been made near the resolution limit and face 
challenges in separating the protoplanet detection from the emission of
asymmetric disk features.
For instance, it has been argued that the phase signature on T\,Cha might
trace disk emission from the inner edge of the outer disk (at around 12\,au),
instead of a planetary-mass companion (Olofsson et al.\cite{olo13}).
Observations on the pre-transitional disks V1247\,Ori also revealed
the presence of complex disk asymmetries in the inner regions of the disk
around this object, whose phase signatures can easily be mistaken as
close companions (Kraus et al.\cite{kra13}).

The $\sim 100$-times higher angular resolution provided by PFI will be needed
to separate the protoplanets from possible disk contributions and to resolve the 
circumplanetary accretion disk on scales of 0.2 Hill radii (0.1\,mas for a Jupiter-mass
planet at 1\,au; e.g.\ Ayliffe \& Bate\cite{ayl09}), enabling us to probe the 
planet formation process over a wide range of spatial scales (Figure~\ref{fig:sketch}).

A major task of this working group will be to identify the optimal line tracers
that are suitable for tracing the circumplanetary accretion disk.  
Intruigingly, recent adaptive optics imaging observations
at visual wavelengths were able to detect the accretion signatures of a low-mass
companion in the pre-transitional disk of HD\,142527 using the H$\alpha$-line tracer (Close et al.\cite{clo14}).
This demonstrates the feasibility for detecting the accretion signatures of low-mass 
objects in pre-main-sequence (PMS) disks.  For PFI we aim to identify further line tracers that 
are suitable for tracing embedded planets.

\subsection{Exoplanetary System Architecture}

Planet population synthesis models aim to reproduce the observed exoplanet
system populations by linking planet formation models 
with models about the dynamical processes of planet-disk interaction,
planet-planet interaction, dust evolution, and accretion physics.
These models depend on a proper knowledge of the relevant processes 
and on the adjustment of a large number of free parameters, which introduces 
major uncertainties.

PFI will change the situation fundamentally, both by providing robust
information about the initial conditions of planet formation and 
by probing the protoplanet distribution during the evolutionary phase
that is most critical for shaping the architecture of exoplanetary systems,
namely the first $\sim 100$\,Myr.
Observing planetary systems during this time interval will allow PFI to
determine where in the disk planets form and how they migrate through 
interaction with the gas-rich disk, providing insights into the mechanisms 
that halt migration, for instance at deadzones, disk truncation points, or 
during the disk dissipation phase (typically at $\sim 10$\,Myr).

Achieving this goal will require detecting planets in a statistically meaningful
sample of systems at different evolutionary phases, e.g.\ of the order of 
100 systems in the classical T\,Tauri/Herbig~Ae/Be phase ($\sim 0.5$\,Myr), 
transitional disk ($\sim 5$\,Myr), 
and early debris disk phase ($\sim 50$\,Myr).
These observations will allow us to construct planet population diagrams
for these different evolutionary phases and to compare them with the
population diagrams for mature exoplanet systems.
Based on state-of-the-art population synthesis models we expect dramatic 
changes during these phases, such as the inward-migration of the 
Hot Jupiters during the first 1-2\,Myr and the ejection of planets
due to dynamical instabilities (e.g.\ Raymond et al.\cite{ray06}).

Another important objective of PFI will be to determine the location of the
``snow line'' for important molecules like water (H$_2$O),
marking the location where these molecules condense to form ice grains.
At the snow line, the density of solid particles in the disk increases abruptly.
This allows planets to form more efficiently beyond the snow line and enables
them to accrete gas from the disk before it dissipates, favouring the
formation of gas giant planets.
The location of the snow line is difficult to predict for different molecules
and changes during the disk evolution.  Therefore, it will be crucial to
measure the location of the snow line directly, for instance using the 
mid-infrared H$_2$O 3.1\,$\mu$m line.
Measuring the distribution of the snow line will allow us
to understand how water is delivered to terrestrial planets like Earth,
where it has been proposed that water can either been produced
through oxidation of atmospheric hydrogen (Ikoma et al.\cite{iko06}) or is
delivered through planetesimals from beyond the snow line (Morbidelli et al.\cite{mor00}).

Given their favourable brightness contrast, PFI will likely focus on giant planets.  
However, the giant planet population is the most critical component in shaping 
the architecture of planetary systems and has also a major impact on the 
formation of terrestrial planets (e.g.\ Morbidelli et al.\cite{mor14}).

\subsection{Late Stages of Planetary System Formation}

The gas- and dust-rich protoplanetary disk dissipates on timescales of 5-10\,Myr 
(Hernandez et al.\cite{her07}), which is also expected to halt type I+II
migration of the embedded protoplanets.  
Smaller amounts of optically thin dust are still observed at the later
debris disk phase.  The dust in these disks is believed to be replenished by the
collision of planetesimals.  
Some debris disks show intruiging ring-like structures that have been 
attributed to the dynamical influence of embedded planets (Kalas et al.\cite{kal05}),
although it has been suggested that these structures might also be shaped by
hydrodynamical instabilities without planets (Lyra \& Kuchner\cite{lyr13}).

The planetary system architecture is still subject to changes during the 
debris disk phase, for instance through planet-planet interaction processes
or through dynamical interaction with the planetesimal disk.
In fact, it is now believed that our solar system also underwent a drastic
reconfiguration at an age of $\sim 700$\,Myr, owing to the migration of
the giant planets due to interaction with the planesimal disk 
and a crossing of Jupiter and Saturn's 1:2 resonance (Tsiganis et al.\cite{tsi05}, Walsh et al.\cite{wal11}).
This reconfiguration might also have triggered the Late Heavy Bombardment
period of the terrestrial planets (Gomez et al.\cite{gom05}).

PFI will be able to image the disk structure and to trace the protoplanets
during the early debris disk phase, while the cooling planets are still relatively bright 
at mid-infrared wavelengths.  Further work will be needed in order to determine
which planet mass and which age range are accessible with PFI.

\subsection{Planet Formation in Multiple Systems}

Another unique application of PFI will be in studying planet formation
in multiple systems.  Several circumbinary planets have already been 
detected in mature systems (e.g.\ Kepler 16, Doyle et al.\cite{doy11}) and 
it has been estimated that more than 1\% of all close binary stars have 
giant planets in nearly coplanar orbits (Welsh et al.\cite{wel12}).
PFI will measure the disk truncation effects that are induced by the companion star
and determine where planet formation is possible in this environment.
The disk measurements will inform us how the tidally cleared gap affect
the migration processes.

\subsection{Star Forming Regions / Target Selection}

This working group will identify the star forming regions that are most suitable
to cover the wide range of evolutionary stages and environments
that we anticipate to probe with PFI.
For this purpose, we will consider not only typical low-mass star forming regions
like Taurus and Ophiuchus, but also high-mass star forming regions like Orion.
Comparing the disk structure in these different environments will inform us
how the planet formation process depends on the mass of the central star
and on environmental factors such as the ambient UV ionization field.

The availability of suitable target star populations on the Northern and Southern Hemisphere
will provide an important input for the TWG, both to identify potential sites
and to formulate the sensitivity requirements for the facility.

\subsection{Secondary Science Cases}

Besides investigating the primary science case, PFI has the potential to revolutionize 
a wide range of other science areas from galactic and extragalactic astronomy.
The unprecedented resolution of $\sim 0.1$\,mas will enable a detailed characterization of 
the fundamental parameters of exoplanet host stars (sizes, limb profile variations)
and to correct for stellar activity features like spots, which limits current exoplanet radial velocity 
and transit studies.  
It will enable direct spectroscopy of Hot Jupiters and the determination of their astrometric orbits.

In galactic astrophysics, some obvious applications could include imaging of photospheric structures,
such as spot patterns on main-sequence stars, the weather patterns on white dwarf stars, 
shock waves on post-AGB stars, and the mass-loss processes around evolved stars.
PFI will be able to study the accretion of matter onto the black hole in X-ray binaries and
to study the circumstellar envelopes and pulsation modes in Cepheids, 
which is essential for a proper calibration of the cosmological distance ladder.

Some unique science goals could also be achieved in extragalactic science, 
for instance by performing detailed imaging of AGN dust tori, by performing spatially
resolved AGN reverberation mapping, by determining dynamical black hole masses 
(through measurement of the rotation profile of the circumnuclear accretion disk), 
and by spatially resolving the multiple images of  gravitational micro-lensing events.

In the initial project phase, these secondary science cases will not drive our design decisions of PFI, 
but they will be taken into account for later refinements of the technical specifications
and for adding further capabilities that do not compromise the primary science mission.

\section{CONCLUSIONS}
\label{sec:conclusions}

The PFI project aims to identify compelling science goals and a 
realistic technical roadmap for taking the next step in exploring the 
universe at high-angular resolution.  
We argue that improved high-angular resolution imaging capabilities will
be inevitable to advance our understanding of the planet formation process.

In this contribution, we outlined our initial thoughts on the key science drivers of PFI,
where our objective is to stimulate the discussion in our science working group 
and in the wider community.
Our top priorities for the next 12-24 months are to define the top-level science requirements
and to present them in a series of white papers.  
This work will provide essential input for the TWG, which will formulate the instrument concept 
and determine feasible facility architectures for meeting the science goals.  

Further information is available on our project website 
\url{http://www.planetformationimager.org}.

\acknowledgments     

We thank the large number of colleagues that volunteered to contribute to our
efforts and that bring our science working group to life.


\bibliography{PFIscience}   
\bibliographystyle{spiebib}   

\end{document}